\documentclass[aps,prl,reprint,superscriptaddress]{revtex4-1}

\usepackage{graphicx}
\usepackage{color}
\usepackage{amsmath}
\usepackage{amsfonts}
\usepackage{amssymb}
\usepackage{dcolumn}
\usepackage{hyperref}
\usepackage{enumerate}
\usepackage{bbold}
\usepackage{cancel}
\def\lb#1{{\textcolor{black}{#1}}}

\begin{document}
  \title{Targeted Recovery as an Effective Strategy against Epidemic Spreading}
\author{L. B\"{o}ttcher}
\email{lucasb@ethz.ch}
\affiliation{ETH Zurich, Wolfgang-Pauli-Strasse 27, CH-8093 Zurich,
Switzerland}
\author{J. S. Andrade Jr.}
\affiliation{
Departamento de F\'isica, Universidade
Federal do Cear\'a, 60451-970 Fortaleza, Cear\'a, Brazil}
\author{H. J. Herrmann}
\affiliation{ETH Zurich, Wolfgang-Pauli-Strasse 27, CH-8093 Zurich,
Switzerland}  
\affiliation{
Departamento de F\'isica, Universidade
Federal do Cear\'a, 60451-970 Fortaleza, Cear\'a, Brazil}
\date{\today}
\begin{abstract}
We propose a targeted intervention protocol where recovery is restricted to individuals that have the least number of infected neighbours. Our recovery strategy is highly efficient on any kind of network, since epidemic outbreaks are minimal when compared to the baseline scenario of spontaneous recovery. In the case of spatially embedded networks, we find that an epidemic stays strongly spatially confined with a characteristic length scale undergoing a random walk. We demonstrate numerically and analytically that this dynamics leads to an epidemic spot with a flat surface structure and a radius that grows linearly with the spreading rate.
\end{abstract}
\maketitle
Modeling epidemic diseases has a long tradition in science \cite{hethcote2000,kermack27}. In recent years the study of epidemic spreading became an attractive research topic in the realm of complex networks leading to a better understanding of the underlying spreading dynamics \cite{gleeson2013,boettcher14,boettcher16,satorras14}. In particular, this progress allowed for the development of efficient countermeasures to contain diseases in our today's highly interconnected world \cite{pastor-satorras2002,Helbing13,brockmann13,colizza2007,suyu2014,Preciado2014}.

More recently, it has been shown that targeted recovery protocols, focusing on influential spreaders \cite{sen14,morone15,nino15}, are able to prevent a system from failure or disease cascades \cite{Gong2015,Muro2016,majdanzic16}. In contrast to targeted recovery strategies, standard models of epidemic spreading assume that nodes, i.e.~the constituents of a network, spontaneously recover after being infected \cite{keeling-rohani2008}. However, nodes located in highly infectious neighborhoods do not only exhibit an enhanced reinfection probability but also accessing these regions might be difficult. Thus, in terms of targeted recovery, one would recover nodes at the boundary of infected regions first. Here we study the latter scenario and find that this strategy is highly efficient in terms of disease containment. The epidemic outbreaks are minimal when compared to the baseline scenario of random intervention. In addition, on spatial networks the dynamics self-organizes to spatially confined epidemics with a characteristic length scale---independently of initial conditions. Similar confined structures emerge in modeling tumor spheroids \cite{sutherland1988,roose2007}, vegetation circles \cite{fernandez2014} and pattern formation in population dynamics \cite{garcia04}. We observe that this confined epidemic spot follows a random walk behavior. Previous studies focused on vaccination strategies to contain a disease \cite{pastor-satorras2002,colizza2007,Preciado2014,suyu2014}. Our results, however, suggest that a targeted recovery approach is also able to drastically reduce the number of infected individuals and to spatially confine an epidemic. Our approach is therefore
of crucial importance in epidemiology when vaccination is not available. This opens completely new avenues of research in the study of spreading dynamics.
\paragraph{Model.}
Our model uses a binary-state dynamics where $N$ nodes are either in a susceptible or in an infected state. In the case of random recovery, infected nodes spontaneously recover at unit rate and susceptible nodes get infected at rate $r$ if at least one neighboring node is infected. Spontaneous transitions from a susceptible to an infected state occur at rate $p$. However, unless specified otherwise, we typically assume $p=0$ and just use this spontaneous infection term to perturb the system. This baseline scenario resembles the critical behavior of the contact process or of susceptible-infected-susceptible (SIS) dynamics \cite{boettcher162,boettcher171,marro05,henkel08}. Our recovery protocol accounts for spontaneous and targeted recovery. In our kinetic Monte Carlo simulations \cite{gillespie76, gillespie77}, the probability of a recovery event to occur is proportional to the total number of infected nodes. With probability $1-\epsilon$ recovery is spontaneous, i.e.~a randomly selected infected node recovers, and with probability $\epsilon$ it is targeted. In the latter case, nodes in the least-infected neighborhoods recover first. We therefore create a list in which infected nodes are sorted according to their fraction of infected neighbours. An infected node with the least fraction of infected neighbours is then randomly selected to recover. The parameter $1-\epsilon$ accounts for spontaneous recovery which might be present in real situations and allows to analyze the stability of the effects in the case where $\epsilon=1$ by perturbing the system with random recovery events. We first study the dynamics on a spatially embedded network to later on assess the influence of a transition to a random network \cite{buckee2004,danziger15}. Spatial models are widely used in epidemiology to understand the influence of spatial effects \cite{white1995,keeling2000} and many real-world systems exhibit features of random networks \cite{gleeson12}.
\paragraph{Targeted recovery and confined epidemics.}
\begin{figure}
\begin{minipage}{0.35\textwidth}
\centering
\includegraphics[width=\textwidth]{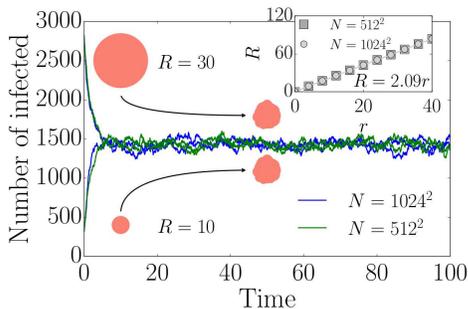}
\end{minipage}
  \caption{\textbf{Time evolution and radius of the confined region.} The number of infected nodes as a function of time for $r=10$, $\epsilon=1$ and two different initial circular spreading seeds. In the stationary state, the epidemic is still confined to a deformed circular spot. The inset shows the spreading radius $R=\sqrt{n_{st}(r) N/\pi}$ of the confined region as a function of the spreading rate $r$. All simulations have been performed on square lattices with $N$ nodes and a circular spreading seed.} 
 \label{fig:dynamics_and_radius}
\end{figure}
We start with the case where nodes in least-infected neighborhoods always recover first, i.e.~$\epsilon\rightarrow 1$. In this case, some nodes in highly infected neighborhoods cannot recover although we assume that the probability of a recovery process is proportional to the population of infected nodes. We account for random recovery events in the subsequent paragraphs thus justifying the latter proportionality. As we see later on, properties such as the formation of confined epidemic spots that occur for $\epsilon\rightarrow 1$ are still observable even when a substantial amount of infected nodes recovers spontaneously. For now we consider a spatially embedded network with a spreading seed made of a circular region of infected nodes. In Fig.~\ref{fig:dynamics_and_radius} a typical time-evolution of the number of infected nodes is presented, cf.~\href{https://vimeo.com/192434219}{video 1}. Independent of the initial size of the circular spreading seed, the dynamics converges to the same stationary number of infected nodes. In the stationary state, the epidemic stays confined inside a deformed circular spot of finite size that grows with infection rate $r$. This effect is shown in the inset of Fig.~\ref{fig:dynamics_and_radius}. Independent of the system size, we observe a linear dependence of the spreading radius $R=\sqrt{n_{st}(r) N/\pi}$ on the spreading rate $r$. Here $n_{st}(r)$ defines the fraction of infected nodes in the stationary state as a function of $r$. The linear dependence is a consequence of the following approximated growth dynamics neglecting any stochasticity:
\begin{equation}
\dot{R}(t)=r \frac{2 \pi R(t)}{\pi R^2(t)+2 \pi R(t)}-\frac{\pi R^2(t)}{\pi R^2(t)+2 \pi R(t)},
\label{eq:radius_circle}
\end{equation}
where $R$ is the radius of the spreading area. The first term accounts for the fact that spreading occurs at the boundary of the circular region whereas the second term describes recovery. Recovery is still proportional to the total number of infected nodes ($\pi R^2(t)$), however, just occurs at the boundary in the limit of $\epsilon\rightarrow 1$ and thus reduces the radius. In agreement with the inset of Fig.~\ref{fig:dynamics_and_radius}, this description explains the linear growth of the radius with the infection rate $r$, i.e.~$R(r)=2 r$ in the stationary state.

But are the confined stationary states a mere consequence of the single initial circular spreading seed or is it possible to observe this effect for multiple and randomly distributed spreading seeds as well? We study the influence of small perturbations ($p>0$) leading to a random configuration of spreading seeds instead of, as before, assuming one circular spreading seed, cf.~\href{https://vimeo.com/192434254}{video 2}. In the \emph{Supplemental Material}, we illustrate that even in this case, the dynamics still approaches a stationary state in which the epidemic stays confined. Up to fluctuations due to spontaneous infections ($p>0$), different initial spreading seeds merge at some point and others vanish spontaneously so that at the end only one confined spot remains \footnote{See the \emph{Supplemental Material} for more information on random initial conditions.}.

\paragraph{Diffusive properties.}
\begin{figure}
\begin{minipage}{0.35\textwidth}
\centering
\includegraphics[width=\textwidth]{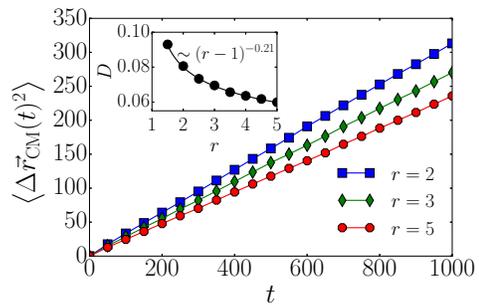}
\end{minipage}
  \caption{\textbf{Random motion of confined epidemics.} The mean-square displacement $\left\langle \Delta \vec{r}_{\mathrm{CM}}(t)^2\right\rangle$ as a function of time for different spreading rates $r$. The linear growth of $\left\langle \Delta \vec{r}_{\mathrm{CM}}(t)^2\right\rangle$ suggests that the confined epidemic performs a random walk. The inset shows the diffusion constant for different spreading rates. The number of samples is $M=2\times 10^4$.} 
\label{fig:msd}
\end{figure}

\begin{figure}
\begin{minipage}{0.35\textwidth}
\centering
\includegraphics[width=\textwidth]{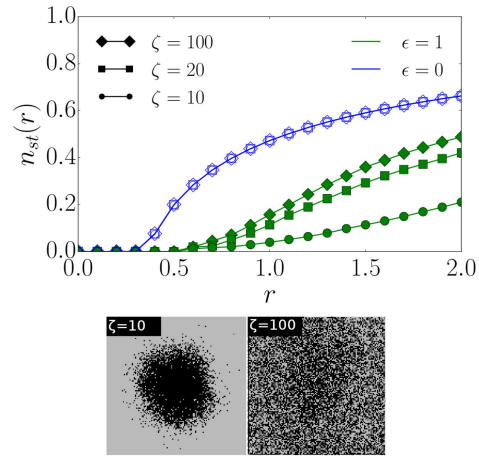}
\end{minipage}
  \caption{\textbf{Transition from a lattice to a random network.} (upper panel) The dependence of $n_{st}(r)$ on $r$ on a network with $N=128\times 128$ nodes for different values of $\zeta$. Targeted recovery with $\epsilon=1$ (green filled markers) outperforms random interventions where $\epsilon=0$ (blue hollow markers) for all values of $\zeta$ since a smaller fraction of nodes is infected in the stationary state. The curves for $\epsilon=0$ and different $\zeta$ lie on top of each other. (lower panel) Snapshots at $t=10$ for $\zeta=10,~100$, $r=2$ and an initial spreading seed with a radius of 30 on a network with $N=128\times 128$ nodes. Black lattice sites are infected.} 
 \label{fig:zeta_lattice}
\end{figure}

The confined circular region moves over the spatially embedded system since its boundaries are always subject to deformations due to infection and recovery events. We define the mean-square displacement of the confined region as
\begin{equation}
\left\langle \Delta \vec{r}_{\mathrm{CM}}(t)^2\right\rangle = \frac{1}{M} \sum_{i=1}^{M} \left(\vec{r}_{\mathrm{CM}}^{\,i}(t)-\vec{r}_{0}^{\,i}\right)^2,
\end{equation}
where $M$ is the number of samples, $\vec{r}_{\mathrm{CM}}(t)$ is the center of mass at time $t$ and $\vec{r}_{0}$ denotes the center of mass at time $t=0$.
To account for periodic boundaries, we applied the algorithm proposed in Ref.~\cite{bai2008}. We illustrate the mean-square displacement for different spreading rates in Fig.~\ref{fig:msd}. The linear dependence of $\left\langle \Delta \vec{r}_{\mathrm{CM}}(t)^2\right\rangle$ on $t$ suggests a diffusive behavior as typical for a random walk \cite{einstein1905}. The inset of Fig.~\ref{fig:msd} shows the diffusion constant $D$ defined by the relation $\left\langle \Delta \vec{r}_{\mathrm{CM}}(t)^2\right\rangle = 4 D t$ for different spreading rates $r$. \lb{Infection and recovery events correspond to center of mass shifts of size $1/R$ and $-1/R$, respectively. In the case of large clusters, i.e.~for a large value of the rate $r$, the diffusion constant is proportional to the squared center of mass shift and to the number of such shifts per unit time $N$, i.e.~$D\sim N R^{-2}$. The number of shifts per unit time is proportional to $R^2$ and thus $D$ is constant. What is the value of this constant? For large clusters the infection and recovery events are uniformly distributed around the perimeter. As a consequence of this uniform distribution, one would expect to see a non-moving cluster characterized by $D\rightarrow 0$ when the size of the cluster goes to infinity in accordance with the observed behavior in Fig.~\ref{fig:msd}. Small confined epidemic clusters do not exhibit a uniform distribution of infection and recovery events due to their finite size and one finds a diffusion constant different from zero as described by the following fit $D(r)\sim (r-1)^{-\alpha}$ with $\alpha=0.21(1)$.} The functional form of $D(r)$ suggests the existence of a threshold $r_{c}(\epsilon=1)=1$. In the case of spreading rates smaller than $r_{c}(\epsilon=1)$, the confined epidemic dies out. Even for values of $r$ slightly larger than 1, the infected spot might disappear due to statistical fluctuations. The threshold value is also implicitly contained in Eq.~(1) since $R$ would only take stationary values smaller than 2 for $r<r_c(\epsilon=1)$. So the dynamics would end up in the absorbing state with zero infected nodes.
\paragraph{Surface roughness.}
\begin{figure}
\begin{minipage}{0.35\textwidth}
\centering
\includegraphics[width=\textwidth]{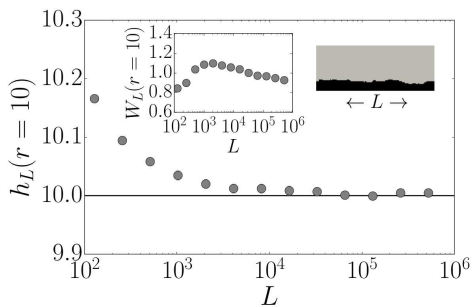}
\end{minipage}
  \caption{\textbf{Surface height and width of the confined epidemic.} The surface height $h_L(r=10)$ for $r=10$ as a function of the linear system size $L\in\{2^6,2^7,\dots,2^{19}\}$ of a square lattice. For the smallest system with $L=64$, the number of samples is $10^4$. As the system size doubles, the number of samples is divided by 2. A typical rectangular interface structure is shown in the upper right inset. Black lattice sites are infected. The width $W_L(r=10)$ as a function of $L$ is shown in the upper left inset. Error bars are smaller than the marker sizes. }
 \label{fig:roughness}
\end{figure}
\begin{figure}
\begin{minipage}{0.35\textwidth}
\centering
\includegraphics[width=\textwidth]{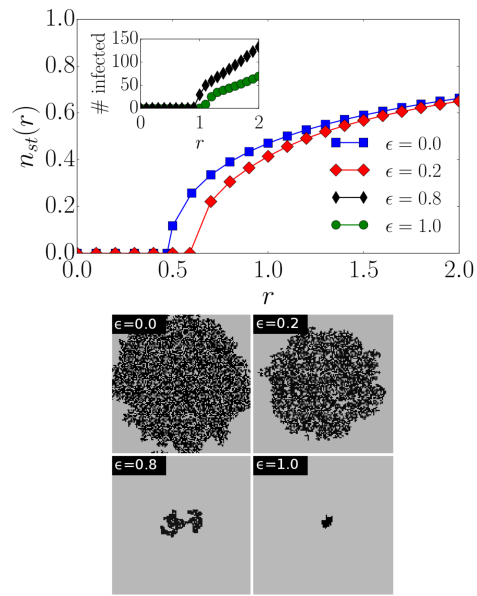}
\end{minipage}
  \caption{\textbf{Influence of spontaneous recovery.} (upper panel) The fraction of infected nodes in the stationary state $n_{st}(r)$ as a function of $r$ for different values of $\epsilon$. The inset shows the number of infected nodes for $\epsilon=0.8$ and $\epsilon=1$ (averaged over 500 samples). Two different presentations have been used since the number of infected nodes is only proportional to the system size for small values of $\epsilon$. Otherwise the number of infected nodes does not depend on the system size. All simulations have been performed on a square lattice with $N=1024\times 1024$ nodes. (lower panel) Snapshots at $t=10$ for different $\epsilon$, $r=2$ and an initial spreading seed with a radius of 30 on a square lattice with $N=128\times 128$ nodes. Black lattice sites correspond to infected ones.} 
 \label{fig:phase_space}
\end{figure}
As illustrated in Fig.~\ref{fig:dynamics_and_radius} and Fig.~S1 in the \emph{Supplemental Material}, the confined epidemic exhibits a characteristic interface separating infected and susceptible nodes. We analyze the surface roughness in terms of its width and define the overall stationary width of a spatially embedded network with linear dimension $L$ and a spreading rate $r$ \cite{interface2011},
\begin{equation}
W_L(r)=\sqrt{\left\langle \left[ h(x)-\langle h \rangle \right]^2\right\rangle},
\end{equation}
where $h(x)$ denotes the interface height. In order to form a stable flat interface, we used as initial condition a line of infected nodes at the bottom of a lattice, cf.~upper right inset in Fig.~\ref{fig:roughness}. We then set the spreading rate to $r=10$ and study the height $h_L(r=10)$ as well as the width $W_L(r=10)$ as a function of $L$. We clearly see in Fig.~\ref{fig:roughness} that the height approaches the analytical stationary value $h_L(r=10)=10$ as $L$ increases. The value $h(r=10)=10$ or more generally $h(r)=r$ corresponds to the stationary state of Eq.~\eqref{eq:radius_circle} by assuming a rectangular spreading regime instead of a circular one. The width as a measure of the standard deviation of the surface height decreases for large values of $L$ as $h_L$ approaches its analytical value, as shown in the upper left inset of Fig.~\ref{fig:roughness}. For small system sizes we observe deviations from this expected behavior since recovery typically occurs at kinks which are moving through the system. The smaller the system the more often the kink will travel through it in a fixed time. This behavior is different from the one typically observed for the growth of interfaces \cite{family1985,interface2011}. Summarizing, we find that the model's dynamics leads to confined spatial structures with a flat interface.
\paragraph{Influence of spontaneous recovery.}
Until this point we have only discussed the properties of a targeted intervention strategy where nodes in least-infected neighborhoods recover first. But how is spontaneous recovery, i.e.~an $\epsilon$ smaller than 1, going to influence the spreading dynamics? As shown in Fig.~\ref{fig:phase_space} (lower panel), confined epidemics still govern the spatial structure of the spreading dynamics for a value of $\epsilon=0.8$ where twenty percent of the recovery events occur randomly. However, as $\epsilon$ approaches zero, the infected nodes are not confined anymore. We also see the influence of $\epsilon$ by comparing the corresponding phase transitions in Fig.~\ref{fig:phase_space} (upper panel). We discuss the dependence of $n_{st}$ on $\epsilon$ in the \emph{Supplemental Material}. A purely random recovery strategy leads to a relatively low critical spreading rate $r_c(\epsilon=0)=0.47(1)$ \cite{boettcher162} when compared to strategies with $\epsilon>0$. This implies that purely random recovery admits non-zero stationary proportions of infected nodes for spreading rates which might be too low to cause any outbreak for $\epsilon>0$. Thus, recovery of nodes in least-infected neighborhoods is very efficient since it reduces the number of infected nodes drastically. This effect is still observable under large perturbations, i.e.~for small values of $\epsilon$, corresponding to a large proportion of spontaneous recovery events. Similar results have been found for random geometric graphs as shown in the \emph{Supplemental Material}.
\paragraph{Transition from the square lattice to a random network.}
The data presented in Fig.~\ref{fig:phase_space} suggest that epidemic transitions on a spatially embedded system are less severe for any $\epsilon > 0$ when compared to purely random recovery. In order to examine the influence of the underlying network structure on our dynamics, we analyze the transition from a spatial network to a random network \cite{buckee2004,danziger15}. We study a spatially embedded network with degree $k=4$ where long-range links replace randomly chosen nearest-neighbor links \cite{danziger15}. The lengths $l$ of the long-range links are distributed according to $P(l)\sim \exp\left(-l/\zeta\right)$, where $\zeta$ defines the characteristic link length. As $\zeta\rightarrow 0$, a square lattice is recovered, whereas the limit $\zeta\rightarrow \infty$ corresponds to a regular random graph since all link lengths are equally likely.
In Fig.~\ref{fig:zeta_lattice} (upper panel), we illustrate the influence of different values of $\zeta$ on the transitions of the targeted recovery strategy ($\epsilon=1$) in comparison to random recovery ($\epsilon=0$). Targeted recovery outperforms random intervention for all values of $\zeta$ in terms of the number of infected nodes. In the \emph{Supplemental Material}, we show that targeted intervention tends towards random recovery for large values of $r$ and for regular random graphs with a large degree. Our strategy substantially improves the control of an epidemic when applied to random networks.
In Fig.~\ref{fig:zeta_lattice} (lower panel), we observe that the epidemic becomes less confined for $\epsilon=1$ and $\zeta=10$ compared to the situation where $\zeta = 0$ in Fig.~\ref{fig:phase_space} (lower panel). In the case of $\zeta=100$, long-range random connections lead to a loss of the spatial structure destroying the confinement.
\paragraph{Concluding remarks.}
We have studied a disease propagation model accounting for targeted recovery where nodes recover first when they are located in least-infected neighborhoods. Surprisingly, this dynamics self-organizes in spatially confined epidemics resembling deformed circular spots of flat surface that perform a random walk on spatially embedded networks. This effect is still observable under large perturbations of random recovery events. Spatially confining an epidemic is of great importance for controlling diseases. Our targeted recovery protocol outperforms random interventions in terms of the number of infected nodes on networks reaching from spatially embedded systems to random ones. 
The results in Refs.~\cite{pastor-satorras2002,suyu2014} suggest that an optimized vaccination strategy is able to reduce the infection prevalence of a SIS model up to 80 \% compared to random vaccination. In our model, we find that targeted recovery leads to vanishingly small outbreaks on spatial networks and reductions of the disease prevalence up to 50 \% on random networks for the studied parameters. It would be desirable to compare the efficiency of our approach with an optimal recovery protocol for general topologies. However, to the best of our knowledge, a general framework which describes an optimal recovery procedure for standard disease models on a general topology has not been studied yet.
\section*{Acknowledgments}
We acknowledge financial support from the ERC Advanced grant number FP7-319968 FlowCCS of the European Research Council, from the Brazilian agencies CNPq, CAPES and FUNCAP, and from the National Institute of Science and Technology for Complex Systems (INCT-SC) in Brazil.
\section*{Author contribution statement}
L.B. carried out computational simulations, analytical calculations and wrote the first draft of the manuscript. L.B., J.S.A. and H.J.H. contributed equally to the ideas, the interpretation and the presentation. 
\section*{Additional information}
\textbf{Competing financial interests} The author(s) declare no competing financial interests.
\end{document}